\documentclass[draft]{agujournal2019_cranmer}
\usepackage{url}
\usepackage{soul}
\usepackage{lineno}
\usepackage[finalnew]{trackchanges} 

\draftfalse
\journalname{JGR: Space Physics}

\begin{document}

\title{Electron Heat Flux in the Solar Wind: Generalized Approaches
to Fluid Transport with a Variety of Skewed Velocity Distributions}

\authors{Steven R. Cranmer\affil{1} and Avery J. Schiff\affil{1,2}}

\affiliation{1}{Department of Astrophysical and Planetary Sciences,
Laboratory for Atmospheric and Space Physics,
University of Colorado, Boulder, CO 80309, USA}

\vspace*{0.10in}
\affiliation{2}{Myriad Genetics, Salt Lake City, UT 84108, USA}

\correspondingauthor{Steven R. Cranmer}{steven.cranmer@colorado.edu}

\begin{keypoints}
\item
Understanding coronal heating and solar wind acceleration requires an
accurate description of heat conduction beyond Spitzer-H\"{a}rm theory.
\item
Fluid-moment models of parallel electron heat conduction are constructed
using skewed, positive-definite velocity distributions.
\item
Both analytic models and fits to in~situ data can be used to
constrain parameters of closed-form solutions for the collisionless
heat flux.
\end{keypoints}

\begin{abstract}
In the solar corona and solar wind, electron heat conduction is an
important process that transports energy over large distances and helps
determine the spatial variation of temperature.
High-density regions undergoing rapid particle-particle collisions
exhibit a heat flux described well by classical Spitzer-H\"{a}rm theory.
However, much of the heliosphere is closer to a more collisionless state,
and there is no standard description of heat conduction for fluid-based
(e.g., magnetohydrodynamic) models that applies generally.
Some proposed models rely on electron velocity distributions that
exhibit negative values of the phase-space density.
In this paper, we explore how positive-definite velocity distributions
can be used in fluid-based conservation equations for the electron heat
flux along magnetic-field lines in the corona and solar wind.
We study both analytic forms of skewed distributions (e.g.,
skew-normal distributions, two-sided bi-Maxwellians, and
constant-collision-time electrostatic solutions) and empirical fits to
measurements of core, halo, and strahl electrons in interplanetary space.
We also present example solutions to a generalized conservation equation
for the heat flux in the solar wind, with some limiting cases found to
resemble known free-streaming approximations.
The resulting values of the electron heat flux vary as a function of
radial distance and Knudsen number in ways that resemble observed data.
We note that this model does not include the effects of kinetic
instabilities (which may impose saturation limits when active), so for
now its regime of applicability is limited to collisionless heat-flux
evolution away from the known instability boundaries in parameter space.
\end{abstract}

\section*{Plain Language Summary}

The solar corona is a layer of hot (million-degree Celsius) gas that
surrounds the Sun and expands into interplanetary space as the
rapidly accelerating solar wind.
Because the gas is so hot, its atoms become ionized on their journey
from the solar surface.
The nuclei of the former atoms are the heaviest of the new particles,
so they carry most of the solar wind's mass.
However, most of its heat is carried by the rapidly-moving electrons
that once orbited those nuclei.
When close to the Sun, electrons transport heat in much the same was as
they do in metal objects on Earth: random collisions cause the heat to
conduct from hot to cold regions.
We study their behavior higher above the Sun's surface, where the
density of particles plummets and random collisions become infrequent.
In these essentially collision-free regions, electrons still carry heat,
but they do it because the solar wind's expansion causes them to
develop unusually shaped probability distributions of velocity.
We explore various mathematical forms for those distributions and show
how they can be used to accurately determine the strength of heat
conduction in future computer simulations of the corona and solar wind.

\section{Introduction}
\label{sec:intro}

The Sun's combined corona/heliosphere system is a hot (i.e., nearly
fully ionized) and expanding plasma composed of mostly hydrogen, some
helium, and a small fraction of heavier elements.
Despite the negligibly small masses of the free electrons in such a plasma,
these particles are important to maintaining overall quasi-neutrality
and a zero-current electrostatic balance.
In~situ measurements have revealed that the electron velocity
distribution function (VDF) often shows four relatively distinct
components:
(1) an approximately Maxwellian {\em core,} often close to being in
thermal equilibrium with the protons,
(2) a higher-energy isotropic {\em halo,} usually with a mild
power-law tail in velocity space,
(3) a magnetic-field-aligned beam or {\em strahl} that indicates
connectivity to the near-Sun corona, and
(4) a much higher-energy isotropic power-law {\em super-halo}
\cite{Pi87,Lin97,Wa15}.
Interactions between these components give rise to macroscopic
phenomena such as heat conduction, and they determine the state of
several key kinetic instabilities \cite<see, e.g.,>[]{Fe75,Ma06,Ve19}.

Particle-particle collisions are infrequent in a plasma as rarefied
as the solar wind.
There have been many useful insights gained from a fully kinetic
``exospheric'' approach to modeling these kinds of systems
\cite<e.g.,>[]{LS73,MV99,Zo04,Ec11}.
It has generally been presumed to be unwise to model collisionless
plasmas as fluids; i.e., to use moment-based equations of hydrodynamics
or magnetohydrodynamics (MHD).
In addition to assuming frequent collisions, many of these sets of
conservation equations are based on the assumption of perfectly
Maxwellian VDFs.
Nevertheless, fluid-based modeling has been quite successful in
predicting many large-scale properties of the corona and solar wind
\cite{P58,Wi88,Lk99,Go18}.
Also, the existence of quasi-Maxwellian cores in the electron, proton,
and heavy-ion VDFs---often observed in regions of space where the
collisionless theory predicts they should not exist---points to some
kind of velocity-space randomization that plays a similar
thermalizing role as particle collisions.

In order to expand the usefulness of fluid-based modeling to regimes
of lower collisionality, it is necessary to build in compatibility with
non-Maxwellian particle distributions.
Some attempts to do this have relied on polynomial-type perturbations
to an underlying Maxwellian \cite<e.g.,>[]{Sc77,CD85,OL96} 
that have the side-effect of making the VDF unphysically negative in
some regions of velocity space.
Thus, in this paper, we begin to explore other ways to use moment-based
conservation equations that are compatible with positive-definite
non-Maxwellian VDFs.
For now, we focus on one specific physical process: skewness in the
electron distribution function that gives rise to a nonzero heat flux.
We derive a generalized third-moment conservation equation that reproduces
classical \citeA{SH53} thermal conductivity in the limit of rapid
collisions, but also can be applied to more collisionless free-streaming
regimes of the solar wind.

Section \ref{sec:cons} of this paper defines several key properties of
the electron VDF---including shape-dependent coefficients related to
the partitioning of energy in third-order and fourth-order moments---and
describes generalized conservation equations for the moments.
Section \ref{sec:vdf} examines a series of different formulations for
the shapes of skewed VDFs that carry nonzero heat flux along
magnetic-field lines.
Section \ref{sec:wind} presents example calculations for the heat flux,
compares with observational data, and shows how some traditional
analytic limits \cite<e.g.,>[]{H74,H76} can be considered as special
cases of the theory developed in this paper.
Section \ref{sec:conc} summarizes the major results of this paper
and discusses some broader implications.

\section{Generalized Conservation Equations}
\label{sec:cons}

In order to devise new ways of modeling the electron heat flux,
we first must describe the shape of the VDF and the behavior of
its kinetic moments as a function of space and time.
Section \ref{sec:cons:mom} sets up the required definitions, and
Section \ref{sec:cons:tube} shows how the fluid-moment
conservation equations reduce to simpler forms when studying
time-steady electron transport along magnetic-field lines.
Section \ref{sec:cons:coll} ties down the terms in these equations
that describe Coulomb collisions and shows how classical \citeA{SH53}
conductivity can be made to apply when those collisions are rapid.

\subsection{Moments in Velocity Space}
\label{sec:cons:mom}

The primary quantities to be modeled are the electron phase-space VDF
$f({\bf v})$ and its velocity-space moments. 
Our notation mainly follows \citeA{Sc77}, but there are a number of
other useful and definitive derivations of these quantities
\cite<e.g.,>[]{Bg65,CC70,BS82,St05,CB06,Hu19a,Hu19b}.
The number density is the zeroth moment (i.e., just the integral
over all velocity space) of the VDF,
\begin{equation}
  n \, = \, \int d^3 {\bf v} \,\, f({\bf v}) \,\, .
\end{equation}
Higher moments are described by weighted averages of the form
\begin{equation}
  \langle \Phi({\bf v}) \rangle \, = \,
  \frac{\int d^3 {\bf v} \,\, f({\bf v}) \, \Phi({\bf v})}
  {\int d^3 {\bf v} \,\, f({\bf v})} 
\end{equation}
where $\Phi({\bf v})$ can be a scalar, vector, or tensor function
of the scalar components of ${\bf v}$.
The first moment, the bulk fluid velocity, is defined
as ${\bf u} = \langle {\bf v} \rangle$.
Most higher moments make use of the velocity measured in the
bulk-flow frame, ${\bf c} = {\bf v} - {\bf u}$, which is
sometimes called the peculiar or random velocity.

We define the second-rank (3~$\times$~3) pressure tensor
and third-rank (3~$\times$~3~$\times$~3) heat-flow tensor, using
dyadic notation, as
\begin{equation}
  {\cal P} \, = \, n m \langle {\bf cc} \rangle
  \,\,\, , \,\,\,\,\,\,\,\,
  {\cal Q} \, = \, n m \langle {\bf ccc} \rangle \,\, ,
\end{equation}
where $m$ refers to the electron mass.
Later we often make use of various contracted forms of these
tensors, such as the scalar pressure
\begin{equation}
  p \, = \, \frac{1}{3} \, \mbox{Tr} ({\cal P})
  \, = \, \frac{1}{3} \sum_{i=1}^3 \, {\cal P}_{ii}
\end{equation}
and the heat-flux vector
\begin{equation}
  {\bf q} \, = \, \frac{1}{2} n m \langle c^2 \, {\bf c} \rangle \,\, .
\end{equation}
Lastly, we define a fourth-order thermodynamic quantity,
for which only a second-rank tensor contraction is needed,
\begin{equation}
  {\cal R} \, = \, \frac{1}{2} n m \langle c^2 \, {\bf cc} \rangle \,\, ,
\end{equation}
and the divergence of this quantity is included in the conservation
equation for heat flux.

Conservation equations for the VDF moments are constructed by taking
appropriate moments of the Boltzmann transport equation.
Because we primarily want to study the effects of VDF skewness on the
evolution of the heat flux ${\bf q}$, we make several simplifications
to isolate these effects from those introduced by other processes:
\begin{enumerate}
\item
Although the general problem of deriving a set of fluid-moment
closure equations usually involves the need to include a range of
source/sink terms (e.g., external forces in the momentum equation
or explicit heating/cooling terms in the thermal energy equation),
we will not include them here.
See, however, Section \ref{sec:wind:ext} for some additional
discussion of these effects.
\item
For now, we will use only the collisionless Vlasov form of the
Boltzmann equation.
The effects of electron--electron Coulomb collisions
\cite<e.g.,>[]{Bg65} will be reincorporated in an approximate way later.
\item
In the absence of VDF skewness, we generally assume the distribution
function is isotropic, such that the pressure tensor is diagonal
and given by $p {\cal I}$ (where ${\cal I}$ is the 3~$\times$~3
identity matrix).
This sets all stress-tensor terms to zero and eliminates viscosity
from the system.
\end{enumerate}
Thus, we follow equation (20) of \citeA{Sc77} and write the
momentum conservation equation as
\begin{equation}
  nm \frac{D{\bf u}}{Dt} + \nabla p \, = \, 0 \,\, ,
  \label{eq:mom_orig}
\end{equation}
where the advective derivative $D/Dt$ is defined as
$\partial / \partial t \, + \, {\bf u} \cdot \nabla$.
The thermal energy conservation equation is given by
\begin{equation}
  \frac{3}{2} \frac{Dp}{Dt} + \frac{3}{2} p (\nabla \cdot {\bf u})
  + {\cal P} : \nabla {\bf u} + \nabla \cdot {\bf q} \, = \, 0
  \label{eq:dpdt_orig}
\end{equation}
where the colon denotes a tensor double dot product.
Finally, the heat-flux conservation equation is
\begin{equation}
  \frac{D{\bf q}}{Dt} + {\bf q} \cdot \nabla {\bf u}
  + {\bf q} (\nabla \cdot {\bf u}) + {\cal Q} : \nabla {\bf u}
  + \nabla \cdot {\cal R} + \left[ \frac{D{\bf u}}{Dt} \cdot
  \left( \frac{5}{2} p {\cal I} \right) \right] \, = \, 0 \,\, .
  \label{eq:dqdt_orig}
\end{equation}
Given the assumptions listed above, these equations are exact
and do not involve any artificial closure truncations.

\subsection{Transport along Magnetic Flux Tubes}
\label{sec:cons:tube}

Many of the vector and tensor terms in the fluid conservation equations
are simplified substantially when restricting the geometry to
one-dimensional trajectories parallel to a large-scale magnetic field.
Anticipating application to the global corona and heliosphere, we
set up a spherical coordinate system in which the magnetic field
is pointed radially away from the origin along the polar axis, and
the only direction along which quantities can vary is $r$.
Thus, all partial derivatives with respect to the locally transverse
coordinates ($\theta$ and $\phi$) are zero.
However, we include the effect of {\em flux-tube expansion;} i.e.,
the nonradial spreading of slightly off-axis field lines centered
around the polar axis.
The properties of field-line coordinate systems like this have
been studied extensively
\cite<see, e.g.,>[]{Za88,GR91,WT93,LL06}.

If the magnitude of the magnetic field varies along the polar axis
as $B(r)$, then the cross-sectional area of an idealized flux tube is
given by the conservation of magnetic flux as $A(r) \propto 1/B(r)$.
This means that vector gradients and divergences can be written as
\begin{equation}
  \nabla f \, = \, \frac{\partial f}{\partial r} \, \hat{\bf e}_r
  \,\,\, , \,\,\,\,\,\,\,
  \nabla \cdot {\bf F} \, = \, \frac{1}{A} \frac{\partial}{\partial r}
  (A \, F_r )  \,\, ,
\end{equation}
where $\hat{\bf e}_r$ is the radial unit vector.
Let us also assume that the only nonzero component of the solar-wind
bulk velocity ${\bf u}$ is its radial (i.e., field-aligned)
component $u_r$, so that, for example,
\begin{equation}
  {\cal P} : \nabla {\bf u}  \, = \,
  {\cal P}_{rr} \frac{\partial u_r}{\partial r} \, + \,
  ( {\cal P}_{\theta\theta} + {\cal P}_{\phi\phi} )
  \frac{u_r}{2A} \frac{\partial A}{\partial r} \,\, .
\end{equation}
Note that the adopted coordinates do not account for the Parker spiral
effect in the heliosphere; i.e., the misalignment of the radial bulk
flow and magnetic field directions in the ecliptic plane.
See, for example, \citeA{Li99} for extensions to the conservation
equations that take this effect into account.

Making use of additional identities for the flux-tube geometry,
the thermal energy equation can be written in these coordinates as
\begin{equation}
  \frac{3}{2} \frac{\partial p}{\partial t} +
  \frac{3}{2} u_r \frac{\partial p}{\partial r} +
  \frac{5}{2} \frac{p}{A} \frac{\partial}{\partial r} ( u_r A ) +
  \frac{1}{A} \frac{\partial}{\partial r} ( q_r A )
  \, = \, 0 \,\, .
  \label{eq:dpdt_v2}
\end{equation}
Also, combining the radial components of equations
(\ref{eq:mom_orig}) and (\ref{eq:dqdt_orig}) gives the field-aligned
component of the heat-flux equation:
\begin{displaymath}
  \frac{\partial q_r}{\partial t} +
  u_r \frac{\partial q_r}{\partial r} +
  q_r \frac{\partial u_r}{\partial r} +
  \frac{q_r}{A} \frac{\partial}{\partial r} ( u_r A ) +
  {\cal Q}_{rrr} \frac{\partial u_r}{\partial r} +
  ({\cal Q}_{r\theta\theta} + {\cal Q}_{r\phi\phi})
    \frac{u_r}{2A} \frac{\partial A}{\partial r} \,\, +
\end{displaymath}
\begin{equation}
  + \,\, \frac{1}{A} \frac{\partial}{\partial r} ( {\cal R}_{rr} A ) -
  ({\cal R}_{\theta\theta} + {\cal R}_{\phi\phi})
    \frac{1}{2A} \frac{\partial A}{\partial r} -
  \frac{5p}{2nm} \frac{\partial p}{\partial r} \, = \, 0 \,\, .
\end{equation}
We can simplify the third-order quantities above by using the
moment definitions to show that
\begin{equation}
  q_r \, = \, \frac{1}{2} \left(
  {\cal Q}_{rrr} + {\cal Q}_{r\theta\theta} +
  {\cal Q}_{r\phi\phi} \right)  \,\, .
  \label{eq:qpartition}
\end{equation}
For the gyrotropic VDFs considered in this paper, we can also
note that ${\cal Q}_{r\theta\theta} = {\cal Q}_{r\phi\phi}$.
We found that it is useful to write equation (\ref{eq:qpartition})
using dimensionless partition fractions $\alpha_{\parallel}$ and
$\alpha_{\perp}$, such that
\begin{equation}
  {\cal Q}_{rrr} \, = \, 2 \alpha_{\parallel} q_r
  \,\,\, , \,\,\,\,\,\,
  {\cal Q}_{r\theta\theta} \, = \,
  {\cal Q}_{r\phi\phi} \, = \, \alpha_{\perp} q_r
\end{equation}
and
\begin{equation}
  \alpha_{\parallel} + \alpha_{\perp} \, = \, 1 \,\,\, .
\end{equation}
Also, we define
\begin{equation}
  {\cal R}_{rr} \, = \, \eta_{\parallel} \, \frac{5p^2}{2mn}
  \,\,\, , \,\,\,\,\,\,
  {\cal R}_{\theta\theta} \, = \,
  {\cal R}_{\phi\phi} \, = \, \eta_{\perp} \, \frac{5p^2}{2mn}
  \label{eq:etadefs}
\end{equation}
and we anticipate the results of Section \ref{sec:vdf} to note that,
in many cases, $\eta_{\parallel} \approx \eta_{\perp} \approx 1$.

We use notation defined above, together with the addition of a
schematic collision term on the right-hand side, to rewrite the
heat-flux conservation equation as
\begin{displaymath}
  \frac{\partial q_r}{\partial t} +
  u_r \frac{\partial q_r}{\partial r} +
  (1 + \alpha_{\parallel}) 2 q_r \frac{\partial u_r}{\partial r} +
  (1 + \alpha_{\perp}) \frac{q_r u_r}{L_{\rm A}}
  \,\, +
\end{displaymath}
\begin{equation}
  + \,\, \frac{5}{2} (\eta_{\parallel} - \eta_{\perp})
    \frac{p^2}{mn L_{\rm A}} +
  \frac{5}{2} (\eta_{\parallel} - 1)
    \frac{p}{mn} \frac{\partial p}{\partial r} +
  \frac{5p \eta_{\parallel}}{2m} \frac{\partial}{\partial r}
    \left( \frac{p}{n} \right) \, = \, -\nu_{\rm eff} q_r \,\, ,
  \label{eq:dqdt_v3}
\end{equation}
where we simplify a few of the terms by defining a flux-tube
expansion scale length
\begin{equation}
  L_{\rm A} \, = \, \frac{A}{\partial A / \partial r}  \,\, .
\end{equation}
Equation (\ref{eq:dqdt_v3}) introduces an
effective electron--electron collision frequency $\nu_{\rm eff}$,
and we derive a more detailed expression for it in
Section \ref{sec:cons:coll}.

There are several ways of solving the above equations for the
electron heat flux.
In the limit of a steady-state system (i.e., in which all time
derivatives are zero), equation (\ref{eq:dqdt_v3}) is an
ordinary differential equation for $q_r(r)$.
\citeA{S20} explored numerical solutions to this equation when the
other moments ($n$, $u_r$, $p$) are specified independently and the
VDF-dependent coefficients ($\alpha_{\parallel}$, $\alpha_{\perp}$,
$\eta_{\parallel}$, $\eta_{\perp}$) are given by the Grad-model
values discussed in Section \ref{sec:vdf:grad} below.
Here, we follow \citeA{CD85}, who showed how equation (\ref{eq:dpdt_v2})
can be solved for $\partial q_r / \partial r$ and substituted into
equation (\ref{eq:dqdt_v3}) to provide a direct analytic solution
for $q_r$.
Using the notation developed above, we express this solution as
\begin{equation}
  q_r \, = \, \frac{q_{\rm num}}{q_{\rm den}} \,\, ,
  \label{eq:qnumden}
\end{equation}
with
\begin{equation}
  q_{\rm num} = \frac{3u_r^2}{5} \frac{\partial p}{\partial r} +
  \frac{u_r^2 p}{L_{\rm A}} + u_r p \frac{\partial u_r}{\partial r} +
  \frac{(\eta_{\perp} - \eta_{\parallel}) p^2}{mn L_{\rm A}} +
  (1 - \eta_{\parallel}) \frac{p}{mn} \frac{\partial p}{\partial r} -
  \eta_{\parallel} \frac{p}{m} \frac{\partial}{\partial r}
  \! \left( \frac{p}{n} \right)
  \label{eq:qnum}
\end{equation}
\begin{equation}
  q_{\rm den} \, = \, \frac{4}{5} (1 + \alpha_{\parallel})
    \frac{\partial u_r}{\partial r} +
  \frac{2\alpha_{\perp} u_r}{5 L_{\rm A}} + \frac{2}{5} \nu_{\rm eff}
  \,\, .
  \label{eq:qden}
\end{equation}
Once we compute representative values for some of the higher-moment
constants (e.g., $\alpha_{\parallel}$, $\alpha_{\perp}$,
$\eta_{\parallel}$, $\eta_{\perp}$) in Section \ref{sec:vdf}, we
present a range of example solutions for $q_r$ in Section \ref{sec:wind}.

\subsection{Incorporating Coulomb Collisions}
\label{sec:cons:coll}

The magnitude of the effective collision rate $\nu_{\rm eff}$ can be
estimated by referring to the classical \cite{SH53,Bg65} results for a
strongly collisional and fully ionized plasma.
That case is written as
\begin{equation}
  q_r \, = \, - \frac{\kappa_0 T^{5/2}}{\ln \Lambda} \,
  \frac{\partial T}{\partial r} \,\, ,
  \label{eq:qSH}
\end{equation}
where $\kappa_0 = 1.84 \times 10^{-10}$ W m$^{-1}$ K$^{-7/2}$
and $\ln\Lambda$ is the Coulomb logarithm.
We also assume the isotropic ideal-gas law $p = n k_{\rm B} T$,
where $k_{\rm B}$ is Boltzmann's constant.
The above expression can be shown to be equivalent to
equations (\ref{eq:qnumden})--(\ref{eq:qden}) in the limit of strong
collisions, no outflow ($u_r = 0$), and relatively weak skewness
($\eta_{\parallel} = \eta_{\perp} = 1$).
In that case, only the final terms in each of
equations (\ref{eq:qnum}) and (\ref{eq:qden}) survive, and the heat
flux is given by
\begin{equation}
  q_r \, \approx \, - \frac{5 p}{2 m \nu_{\rm eff}}
  \frac{\partial}{\partial r} \left( \frac{p}{n} \right)
  \, = \, - \frac{5 n k_{\rm B}^2 T}{2 m \nu_{\rm eff}} \,
  \frac{\partial T}{\partial r} \,\, ,
  \label{eq:qcolimit}
\end{equation}
and we emphasize the assumption of $\eta_{\parallel} = 1$ here.
A generic electron--electron collision rate can be written as
\begin{equation}
  \nu_{\rm eff} \, = \, \frac{\xi \, e^4 \, n \, \ln\Lambda}
  {16\pi \varepsilon_0^2 \, m_e^{1/2} (k_{\rm B} T)^{3/2}}  \,\, ,
  \label{eq:nucoll}
\end{equation}
where $e$ is the electron charge, $\varepsilon_0$ is the vacuum
electric permittivity, and $\xi$ is a dimensionless factor.
For specific models of the skewed VDF, such as those discussed
below in Section \ref{sec:vdf}, it is possible to evaluate the
Boltzmann collision term to determine self-consistent expressions
for $\xi$ \cite<see, for example,>[]{Ki04}.
However, in this paper we take a shortcut that presumes the
\citeA{SH53} result is always valid in the limit of rapid collisions,
and thus we aim to determine the value of $\xi$ required to make
that happen.
Thus, incorporating equation (\ref{eq:nucoll}) into
equation (\ref{eq:qcolimit}) and comparing with equation (\ref{eq:qSH}),
we notice the following identity,
\begin{equation}
  \kappa_0 \, = \, \frac{40\pi \varepsilon_0^2 \, k_{\rm B}^{7/2}}
  {\xi \, m_e^{1/2} e^4}
  \, = \, \frac{1.5318 \times 10^{-10} \,\,
  \mbox{W m$^{-1}$ K$^{-7/2}$}}{\xi}
  \,\, ,
\end{equation}
and a specific value of $\xi = 0.8325$ would be required to ensure
equality with the \citeA{SH53} conductivity.
Additional complications that arise for $\eta_{\parallel} \neq 1$
are discussed in Section \ref{sec:wind:scaling} below.

\section{Skewed Velocity Distribution Functions}
\label{sec:vdf}

The goal of this paper is to show how equations
(\ref{eq:qnumden})--(\ref{eq:qden}) can be used to compute values of
the electron heat flux that apply in regions bridging the collisional
and collisionless regimes.
To do that, values for the higher-moment VDF parameters (e.g.,
$\alpha_{\parallel}$, $\alpha_{\perp}$,
$\eta_{\parallel}$, $\eta_{\perp}$) must be known.
Here, we evaluate those parameters for a range of different
analytic expressions of skewed VDFs.
These distributions are all assumed to be gyrotropic; i.e., they are
symmetric along all velocity-space axes transverse to the magnetic
field, and they depend explicitly only on velocity components
$v_{\parallel} = v_r$ and $v_{\perp} = (v_{\theta}^2 + v_{\phi}^2)^{1/2}$.

\subsection{Classical Non-Skewed Basis Functions}
\label{sec:vdf:max}

Because we generally treat skewness as a perturbation to an
underlying equilibrium distribution, the properties of those
non-skewed VDFs should first be specified.
Note, however, that other approaches have been adopted.
For example, \citeA{MN73} learned a lot about the physics of heat flux
by using simple piecewise-constant VDFs.
However, for applications to the solar wind, we find that the
Maxwell-Boltzmann distribution---together with two of its most common
non-skewed generalizations---is a preferable foundation to build upon.

The first generalization is the bi-Maxwellian VDF:
\begin{equation}
  f({\bf v}) \, = \, \frac{n}{\pi^{3/2} w_{\parallel} w_{\perp}^2}
  \exp \left[ -\frac{(v_{\parallel} - u_{\parallel})^2}{w_{\parallel}^2}
  -\frac{v_{\perp}^2}{w_{\perp}^2} \right] \,\, ,
  \label{eq:bimax}
\end{equation}
where the thermal speeds are defined as
$w_{\parallel}^2 = 2k_{\rm B}T_{\parallel}/m$ and
$w_{\perp}^2 = 2k_{\rm B}T_{\perp}/m$.
The distribution is normalized such that both $n$ and $u_{\parallel}$
are the results for the zeroth and first moments.
In general, the pressure tensor is not isotropic, since
\begin{equation}
  {\cal P}_{rr} \, = \, \frac{1}{2} mnw_{\parallel}^2
  \,\,\, , \,\,\,\,\,\,
  {\cal P}_{\theta\theta} \, = \,
  {\cal P}_{\phi\phi} \, = \, \frac{1}{2} mnw_{\perp}^2 \,\, ,
\end{equation}
and the isotropic Maxwellian distribution is given when
$w_{\parallel} = w_{\perp} \equiv w$.
In that limiting case,
\begin{equation}
  p \, = \, n k_{\rm B} \left( \frac{T_{\parallel} + 2T_{\perp}}{3}
  \right) \, = \, n k_{\rm B} T 
  \, = \, \frac{1}{2} nmw^2 \,\, .
\end{equation}
Because equation (\ref{eq:bimax}) has no skewness, the corresponding
heat flux vector ${\bf q}$ is identically zero and the
$\alpha_{\parallel}$ and $\alpha_{\perp}$ coefficients are undefined.
However, we can use the second-order moments to define a saturated
heat flux $q_0$, i.e., the flux corresponding to a plasma that
transports its own thermal energy density at the local thermal speed.
One traditional definition for this quantity is
\begin{equation}
  q_0 \, = \, \frac{3}{2} nk_{\rm B}T \sqrt{\frac{2k_{\rm B}T}{m}}
\end{equation}
though other versions---essentially the above expression multiplied by
other order-unity constants---have been proposed
\cite<see, e.g.,>[]{P64,CM77,SL79}.
For a bi-Maxwellian distribution we assume the transport is taking place
along the $r$ direction (i.e., parallel to the field), so that
\begin{equation}
  q_0 \, = \, \frac{mnw_{\parallel}}{2}
  \left( \frac{w_{\parallel}^2}{2} + w_{\perp}^2 \right) \,\, .
\end{equation}
Also, the fourth-order moments are specified by
\begin{equation}
  {\cal R}_{rr} \, = \, mn \left( \frac{3}{8} w_{\parallel}^4 +
  \frac{1}{4} w_{\parallel}^2 w_{\perp}^2 \right)
  \,\,\, , \,\,\,\,\,\,
  {\cal R}_{\theta\theta} \, = \, {\cal R}_{\phi\phi} \, = \,
  mn \left( \frac{1}{8} w_{\parallel}^2 w_{\perp}^2 +
  \frac{1}{2} w_{\perp}^4 \right) \,\, .
\end{equation}
In the isotropic limit, these expressions combine with
equation (\ref{eq:etadefs}) to provide
\begin{equation}
  \eta_{\parallel} \, = \, \eta_{\perp} \, = \, 1 \,\, .
\end{equation}

Another way to generalize the Maxwellian distribution is to add
a suprathermal power-law tail.
In space physics, a common way to parameterize this kind of VDF
is the kappa \cite{Ob68,Va68,LM13}
or bi-kappa \cite<e.g.,>[]{ST92} distribution.
Here, we discuss only the isotropic kappa distribution, for which
the primary independent variable is the magnitude of the
peculiar velocity $c = |{\bf v}-{\bf u}|$, with
\begin{equation}
  f({\bf v}) \, = \, \frac{n \, \Gamma(\kappa + 1)}
  {\pi^{3/2} w^3 (\kappa - 3/2)^{3/2} \Gamma(\kappa - 1/2)}
  \left[ 1 + \frac{c^2}{w^2 (\kappa - 3/2)} \right]^{-1-\kappa}
  \,\, .
\end{equation}
This expression reduces to a Maxwellian distribution in the limit
$\kappa \rightarrow \infty$.
With the form of $f$ given above, the second moment
converges to the usual value ($p = nk_{\rm B} T$) only for
$\kappa > 3/2$ and diverges for smaller values of $\kappa$.
Keeping the fourth moment finite provides a more stringent limit on the
$\kappa$ exponent, with
\begin{equation}
  \eta_{\parallel} \, = \, \eta_{\perp} \, = \,
  \frac{\kappa - 3/2}{\kappa - 5/2}
  \label{eq:kappa4th}
\end{equation}
only being valid for $\kappa > 5/2$.
Note that the existence of suprathermal power-law wings
enhances these quantities over their equivalent Maxwellian values.
For typical values of $\kappa$ of order 3 to 5, the above values
of $\eta_{\parallel}$ and $\eta_{\perp}$ can range between 1.4 and 3.

\subsection{Grad-Type Polynomial Expansions}
\label{sec:vdf:grad}

There are several long-studied ways of perturbing a zeroth-order
non-skewed VDF to be able to solve for transport coefficients such as
the heat flux.
The Chapman--Enskog approach \cite<see, e.g.,>[]{CC70} involves
expanding the distribution function in powers of the Knudsen number
(see Section \ref{sec:wind:num}) and truncating the expansion to
be consistent with a finite number of moment-based fluid equations.
In the \citeA{Grad49} approach, perturbations to the VDF are
expressed as Hermite polynomials, with coefficients determined
by demanding consistency with a finite number of moments.
Here, we focus on Grad-type expansions investigated by \citeA{Sc77}
and \citeA{Ki04}, but note that other related alternatives exist
\cite<see, e.g.,>[]{Wh71,Ch11}.
We must also mention that these expansion-based VDFs all tend to
exhibit a clearly undesirable feature: they require that some regions
of velocity space have $f({\bf v}) < 0$.
This unphysical property is discussed further below.

\citeA{Sc77} developed the classical \citeA{Grad49} expansion, also
known as the eight-moment approximation, whose VDF can be written as
\begin{equation}
  f({\bf v}) \, = \, \frac{n}{\pi^{3/2} w^3} \exp \left(
  - \frac{c^2}{w^2} \right) \left[ 1 -
  \frac{3 q_r c_{\parallel}}{q_0 w}
  \left( 1 - \frac{2 c^2}{5 w^2} \right) \right]
  \,\, .
  \label{eq:vdf_schunk}
\end{equation}
Note that \citeA{DS79} generalized this to an even more general expansion
around a bi-Maxwellian basis function.
In equation (\ref{eq:vdf_schunk}), the saturated heat flux is given by
its isotropic Maxwellian limit, $q_0 = (3/4) mnw^3$.
This distribution has been parameterized so that the zeroth, first,
second, and third moments all work out to their proper values as defined
above, no matter how large is the specified value of $q_r / q_0$.
The components of the heat-flux tensor correspond to values of
\begin{equation}
  \alpha_{\parallel} \, = \, \frac{3}{5}
  \,\,\, , \,\,\,\,\,\,\,
  \alpha_{\perp} \, = \, \frac{2}{5} \,\, .
  \label{eq:alpha_grad}
\end{equation}
Also, this VDF shares the same fourth-order moment behavior
as the corresponding Maxwellian (i.e., the limit $q_r \rightarrow 0$)
in that the ${\cal R}$ tensor is diagonal, with
$\eta_{\parallel} = \eta_{\perp} = 1$.

\citeA{Ki04} noted that the first-order corrections to a Maxwellian
in equation (\ref{eq:vdf_schunk}) do not agree with the expected
velocity-space scalings of \citeA{SH53} theory, so they proposed a
revised version:
\begin{equation}
  f({\bf v}) \, = \, \frac{n}{\pi^{3/2} w^3} \exp \left(
  - \frac{c^2}{w^2} \right) \left[ 1 -
  \frac{6 q_r c_{\parallel} c^2}{5 q_0 w^3}
  \left( 1 - \frac{2 c^2}{7 w^2} \right) \right] \,\, .
  \label{eq:vdf_killie}
\end{equation}
Despite the fact that equation (\ref{eq:vdf_killie}) results in
a significant improvement in accuracy for the Coulomb collision term,
in comparison with equation (\ref{eq:vdf_schunk}),
all of the moments defined in this paper are identical for the
\citeA{Ki04} and \citeA{Sc77} forms of the VDF.

\begin{figure}[!t]
\centering%
\noindent\includegraphics[width=0.93\textwidth]{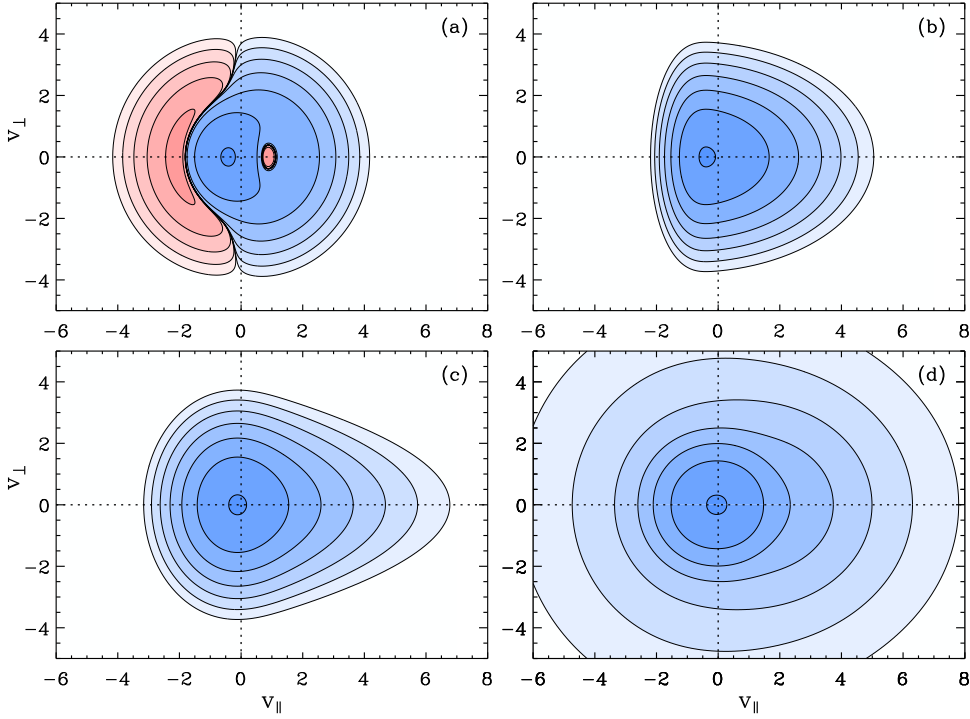}
\caption{Example skewed VDFs, plotted as one-per-decade contours with
$f > 0$ (blue) and $f < 0$ (red), each starting at $\pm$0.9 times the
peak value in each case.  Velocities $v_{\parallel}$ and $v_{\perp}$
are given in units of each distribution's thermal speed $w$, and
have been shifted to put the bulk speed ${\bf u}$ at the origin.
(a) \citeA{Sc77} VDF with $q_r/q_0 = 0.6$.
(b) Skew-normal VDF with $\sigma_s = 3.9$.
(c) \citeA{WW72} VDF with $\sigma_w = 1.4$.
(d) {\em Wind} VDF fit from 16 April 1999, 10:30:27 UT \cite{Wi19a}.}
\label{fig01}
\end{figure}

Figure \ref{fig01}a shows an example VDF, using the \citeA{Sc77} form,
as contours in gyrotropic ($v_{\parallel}$, $v_{\perp}$) coordinates.
Note the existence of multiple regions of velocity space in which $f < 0$.
One way of thinking about these regions is that the conservation of
system-wide quantities like mass, momentum, and energy comes at the
expense of creating a small population of unphysical ``antiparticles.''
\citeA{Sd19,Sd21} discussed several reasons why these kinds of VDFs
should not be used, especially when the magnitudes of the skewness
and heat flux become large.
Thus, we include a discussion of these distributions mainly for
completeness, and we do not recommend basing solutions to
equations (\ref{eq:qnumden})--(\ref{eq:qden}) on their properties.

\subsection{Two-Sided Bi-Maxwellians}
\label{sec:vdf:dimax}

The remaining four subsections of Section \ref{sec:vdf} involve
finding ways of representing skewness in an electron VDF without
creating regions of negative $f({\bf v})$.

One straightforward way to specify a skewed VDF would be to assign
different thermal widths to the two halves of the distribution
corresponding to particles moving in the $+r$ and $-r$ directions.
Something similar to this idea has been used, for kappa distributions,
by \citeA{Lz12}.
Also, studies of rarefied gas flows in the vicinity of a solid, flat
plane traditionally use a similar kind of two-sided distribution
\cite<e.g.,>[]{ES85}.
The basic form of the proposed VDF, written in a frame at rest
for an unskewed distribution, is
\begin{equation}
  f({\bf v}) \, = \, \frac{2n}{\pi^{3/2} w_0^2 (w_1 + w_2)}
  \exp \left( - \frac{v_{\perp}^2}{w_0^2} \right) \times \left\{
  \begin{array}{ll}
   \exp( -v_{\parallel}^2 / w_1^2 ) & \mbox{\, for $v_{\parallel} < 0$,} \\
   \exp( -v_{\parallel}^2 / w_2^2 ) & \mbox{\, for $v_{\parallel} \geq 0$} \\
  \end{array} \right.
  \label{eq:trimax}
\end{equation}
where the three thermal speeds ($w_0$, $w_1$, $w_2$) essentially make
this a ``tri-Maxwellian.''
In general, we utilize $w_2 > w_1$ to produce a positive heat flux
along the parallel axis.
This gives rise to a positive first moment $u_r = (w_2 - w_1)/\sqrt{\pi}$
which must be taken into account when computing
${\bf c} = {\bf v} - {\bf u}$ for the higher moments.
We do not plot an example of this VDF in Figure \ref{fig01} because
its contours look almost the same as the
skew-normal distribution discussed below.
The skewness inherent in equation (\ref{eq:trimax}) can be parameterized
by a dimensionless asymmetry parameter $\sigma_t = (w_2 / w_1) - 1$.
In order for the pressure to be isotropic (i.e., for ${\cal P}_{rr}$
to remain equal to ${\cal P}_{\theta\theta}$ and ${\cal P}_{\phi\phi}$),
we must also set
\begin{equation}
  w_0^2 \, = \, \left[ \frac{(\sigma_t + 1)^3 + 1}{\sigma_t + 2} -
  \frac{2 \sigma_t^2}{\pi} \right] w_1^2  \,\, ,
\end{equation}
and it is clear that for $\sigma_t = 0$, the parameters reduce to
$w_0 = w_1 = w_2$.

\begin{figure}[!t]
\centering%
\noindent\includegraphics[width=0.99\textwidth]{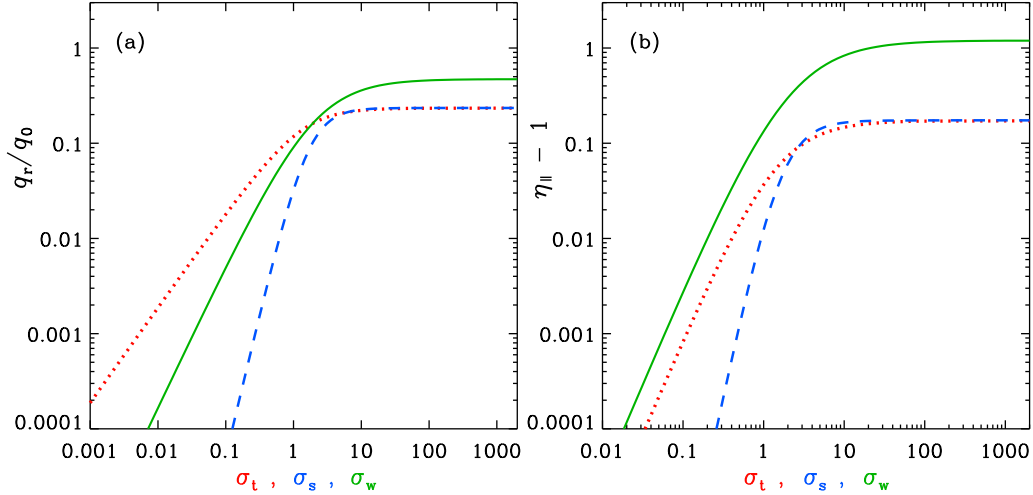}
\caption{(a) Normalized heat flux $q_r / q_0$ plotted as a function
of skewness parameters
$\sigma_t$ (for the two-sided bi-Maxwellian VDF; red dotted curve), and
$\sigma_s$ (for the skew-normal VDF; blue dashed curve), and
$\sigma_w$ (for the BGK \citeauthor{WW72} VDF; green solid curve).
(b) Fourth-order moment coefficient ($\eta_{\parallel}-1$) plotted as
a function of the same skewness parameters shown in panel (a).}
\label{fig02}
\end{figure}

We solved for the zeroth through fourth moments of this two-sided VDF
using a numerical code that simulates the full three-dimensional
velocity space using discrete grids of no less than 1000 points in
$v_r$, $v_{\theta}$, and $v_{\phi}$, with bounds of $\pm 10 w_0$ along
each axis.
Figure \ref{fig02}a shows the resulting parameter dependence of
$q_r / q_0$ as a function of $\sigma_t$.
For small values of the asymmetry parameter, $q_r / q_0$ increases
linearly with $\sigma_t$, then it saturates to a maximum value as
$\sigma_t \rightarrow \infty$.
In this asymptotic limit, the VDF is essentially just one-sided or
hemispherical; i.e., the $v_{\parallel} > 0$ bi-Maxwellian piece is
present, but the $v_{\parallel} < 0$ piece has collapsed to a thin
sheath around $v_{\parallel} \approx 0$.
The limiting value for the heat flux can then be evaluated analytically,
and it is given exactly by
\begin{equation}
  \mbox{max} \left( \frac{q_r}{q_0} \right) \, = \,
  \frac{4-\pi}{3 (\pi - 2)^{3/2}} \, \approx \, 0.234588  \,\, .
\end{equation}
Note that for all values of $\sigma_t$, the heat-flux tensor is
partitioned such that
\begin{equation}
  \alpha_{\parallel} \, = \, 1
  \,\,\, , \,\,\,\,\,\,\,
  \alpha_{\perp} \, = \, 0 \,\, ,
\end{equation}
which is quite different from the Grad-expansion coefficients
given in equation (\ref{eq:alpha_grad}).

For the fourth-order moments, we found that $\eta_{\perp} = 1$ for
all values of $\sigma_t$, but $\eta_{\parallel}$ is an increasing
function of $\sigma_t$.
Figure \ref{fig02}b shows this dependence on the asymmetry parameter.
For low values of $\sigma_t$, $\eta_{\parallel} \approx 1$ as it
should be for a non-skewed Maxwellian.
For high values of $\sigma_t$, $\eta_{\parallel}$ saturates at a value
of about 1.1738.

\subsection{Skew-Normal Distributions}
\label{sec:vdf:gsn}

Originally developed by \citeA{OL76}, the skew-normal distribution is
a generalization of the normal (Gaussian) distribution to nonzero
values of the third moment.
This distribution has been applied in many fields
\cite<see, e.g.,>[]{AC14} and it has been used to model particle
VDFs in systems under the influence of specific types of magnetized
Hamiltonians \cite{Ya15}.

As above, we write the proposed VDF in a frame in which the bulk
flow speed is zero in the limit of no skewness:
\begin{equation}
  f({\bf v}) \, = \, \frac{n}{\pi^{3/2} w_{\parallel} w_{\perp}^2}
  \exp \left( - \frac{v_{\parallel}^2}{w_{\parallel}^2} 
  - \frac{v_{\perp}^2}{w_{\perp}^2} \right) \left[ 1 + \mbox{erf}
  \left( \frac{\sigma_s v_{\parallel}}{w_{\parallel}} \right) \right]
  \,\, ,
\end{equation}
where erf is the Gaussian error function and
$\sigma_s$ is a skewness parameter similar to $\sigma_t$ from
the previous section, in that the limit of $\sigma_s = 0$ represents
zero skewness, and thus zero heat flux.
When $\sigma_s > 0$, there is a nonzero bulk flow speed
\begin{equation}
  u_r \, = \, \frac{\sigma_s w_{\parallel}}{\sqrt{\pi (1 + \sigma_s^2)}}
\end{equation}
and the pressure tensor can be made isotropic if the following
condition on the second moments is imposed:
\begin{equation}
  w_{\perp}^2 \, = \, w_{\parallel}^2 \left[ 1 -
  \frac{2 \sigma_s^2}{\pi (1 + \sigma_s^2)} \right] \,\, .
\end{equation}
Figure \ref{fig01}b shows an example set of contours for this VDF,
once the isotropization has been imposed, and we note that it looks very
similar to the two-sided bi-Maxwellian from Section \ref{sec:vdf:dimax}.

We solved for the higher moments using the same numerical code as
discussed above, and we show in Figure \ref{fig02}a how $q_r / q_0$
depends on $\sigma_s$.
In the low-skewness limit, $q_r / q_0$ increases quite steeply,
in proportion to $\sigma_s^3$.
In the high-skewness limit, the VDF approaches the same shape as the
two-sided bi-Maxwellian, so $q_r / q_0$ approaches the same limiting
value of 0.234588.
For all values of $\sigma_s$, the heat-flux tensor also has the
same partitioning behavior as the VDF from Section \ref{sec:vdf:dimax},
with $\alpha_{\parallel} = 1$ and $\alpha_{\perp} = 0$.
There is also similar behavior for the fourth moments, with
$\eta_{\perp} = 1$ for all values of the skewness, and the same
maximum value of $\eta_{\parallel} = 1.1738$.
Figure \ref{fig02}b shows how the approach to this value differs
from the two-sided bi-Maxwellian case, though.

\subsection{Electrostatic BGK Solutions}
\label{sec:vdf:whealton}

When studying the kinetic properties of a moderately ionized plasma
subject to an electric field, \citeA{WW72} found an elegant
solution to the Boltzmann transport equation.
By making use of the constant collision-time approximation of
\citeA{BGK}, hereafter BGK, they found an analytic solution 
that carries a nonzero heat flux without any regions of negative VDF.
In terms of a dimensionless skewness parameter $\sigma_w$, we can
write this distribution as
\begin{equation}
  f({\bf v}) \, = \, \frac{n \Sigma}{2\pi e w^3}
  \exp \left( - \frac{v_{\perp}^2}{w^2} - \frac{v_{\parallel} \Sigma}{w}
  + \frac{1}{\sigma_w} \right)
  \mbox{erfc} \left[ \frac{1}{\sqrt{\sigma_w}} -
  \frac{\sqrt{\sigma_w}}{2} \left( 1 + \frac{v_{\parallel} \Sigma}{w}
  \right) \right] \,\, ,
\end{equation}
where $\Sigma = \sqrt{(4 + 2\sigma_w)/\sigma_w}$ and
erfc is the complementary error function.
When $\sigma_w \rightarrow 0$, this distribution reduces to an
isotropic Maxwellian.
Applications of this VDF to the solar wind have been proposed by
\citeA{LH97,LH98}, and Figure \ref{fig01}c illustrates its shape.

The above form for $f({\bf v})$ is convenient for analysis, since its
first moment automatically integrates to $u_r = 0$ no matter the value
of $\sigma_w$, and its second-moment pressure-tensor components are
always isotropic.
For the third moment, Figure \ref{fig02}a shows how the numerically
computed values of $q_r/q_0$ approach a constant asymptotic value for
large values of $\sigma_w$, and this value is different than the ones
discussed above.
In this case, the $\sigma_w \rightarrow \infty$ limit produces a
truncated exponential function along the parallel-velocity axis, with
\begin{equation}
  f({\bf v}) \, \approx \, \frac{n\sqrt{2}}{\pi e w^3}
  \exp \left( - \frac{v_{\perp}^2}{w^2} - \frac{v_{\parallel} \sqrt{2}}{w}
  \right) \, H ( v_{\parallel} + w / \sqrt{2} ) \,\, ,
\end{equation}
and the Heaviside step function $H(x)$ is used to indicate that
the asymptotic distribution function is zero for all
$v_{\parallel} < -w/\sqrt{2}$.
For this distribution, $q_r / q_0 = \sqrt{2}/3 \approx 0.4714$, almost
exactly twice the limiting value for the two-sided bi-Maxwellian and
skew-normal VDFs.

For all values of $\sigma_w$, the \citeA{WW72} distribution exhibits
$\alpha_{\parallel} = 1$ and $\alpha_{\perp} = 0$.
This type of third-moment partitioning seems to be ubiquitous for
VDFs that are separable functions of a skewed one-dimensional
distribution in the parallel direction and a Gaussian function in
the perpendicular direction.
The fourth-moment behavior of this VDF is also similar to the
distributions discussed in Sections \ref{sec:vdf:dimax}--\ref{sec:vdf:gsn},
since $\eta_{\perp} = 1$, and $\eta_{\parallel}$
increases monotonically as a function of the skewness parameter
$\sigma_w$ as shown in Figure \ref{fig02}b.
The maximum value of $\eta_{\parallel}$ in this case is 11/5.

\subsection{Multi-Component Fits to Data}
\label{sec:vdf:data}

Analytic parameterizations of VDF skewness can be of great pedagogical
interest, but we also want to ensure that our models resemble the
electron distributions actually observed in the solar wind.
Thus, we sought high-precision functional fits to the core, halo, and
strahl components of electron VDFs measured at specific times.
When the centroids of these components drift relative to one another,
they correspond to nonzero heat fluxes.
These VDFs have been measured, split into components, and fit with
various functions \change{since the 1960s}{for several decades}
\cite<see, e.g.,>[]{Sv09}.
For the purposes of this paper, we focused on publicly available
data, based on statistically robust fitting procedures, that retain
the precision needed to compute third-order and fourth-order moments.
The two sources we used are described as follows:
\begin{enumerate}
\item
\citeA{Wi19a} performed multi-parameter fits to a set of 15,314
electron VDF measurements made at 1~AU, between 1995 and 2000, with
the {\em Wind} spacecraft's 3DP low-energy electrostatic analyzer.
These data were selected for proximity to known interplanetary
shocks, but several statistical features were found to be similar to
known trends in the ambient solar wind \cite<see also>[]{Wi19b,Wi20}.
Out of that entire data set, we identified a subset of 2,840
measurements that had the highest-quality fit-flag ratings and
nonzero values of the relevant number densities.
We used the fit parameters provided in the online data-set of
\citeA{Wi19a} and reconstructed the VDFs using their specified
bi-Maxwellian, bi-kappa, and bi-self-similar fitting functions.
Figure \ref{fig01}d shows an example VDF reconstruction that
corresponds to one of the largest measured values of $q_r / q_0$.
\item
\citeA{Sk21} provided two examples of multicomponent VDF fits made
with data from the {\em Ulysses} SWOOPS (Solar Wind Observations Over
the Poles of the Sun) instrument, from days 288 and 365 of the year 2002.
This was a time period when {\em Ulysses} was in the ambient solar wind
at heliocentric distances between 4.2 and 4.5~AU.
The fitting functions included drifting bi-Maxwellians and
generalized anisotropic $\kappa$ distributions.
\end{enumerate}

We computed full sets of moments for each VDF using a similar numerical
code to the one that produced the results shown in Figure \ref{fig02}.
Because the measured heat fluxes are sometimes oriented in the sunward
direction, and sometimes anti-sunward, we took absolute values of $q_r$
in order to treat both cases on equal footing.
The full range of computed $q_r/q_0$ ratios for the \citeA{Wi19a} data
spanned more than three orders of magnitude, from
$1.02 \times 10^{-4}$ to 0.209.
The mean of the distribution of 2,840 values was 0.0470, and the
corresponding median was 0.0379, with one-standard-deviation bounds
(i.e., 16\% and 84\% percentiles) at values of 0.0139 and 0.0879.
The two \citeA{Sk21} models had large normalized heat flux values
of 0.2003 and 0.3201.
These values appear consistent with other statistical studies of the
electron heat flux at 1~AU \cite<e.g.,>[]{Sa03,Ba13}.

\begin{figure}[!t]
\centering%
\noindent\includegraphics[width=0.99\textwidth]{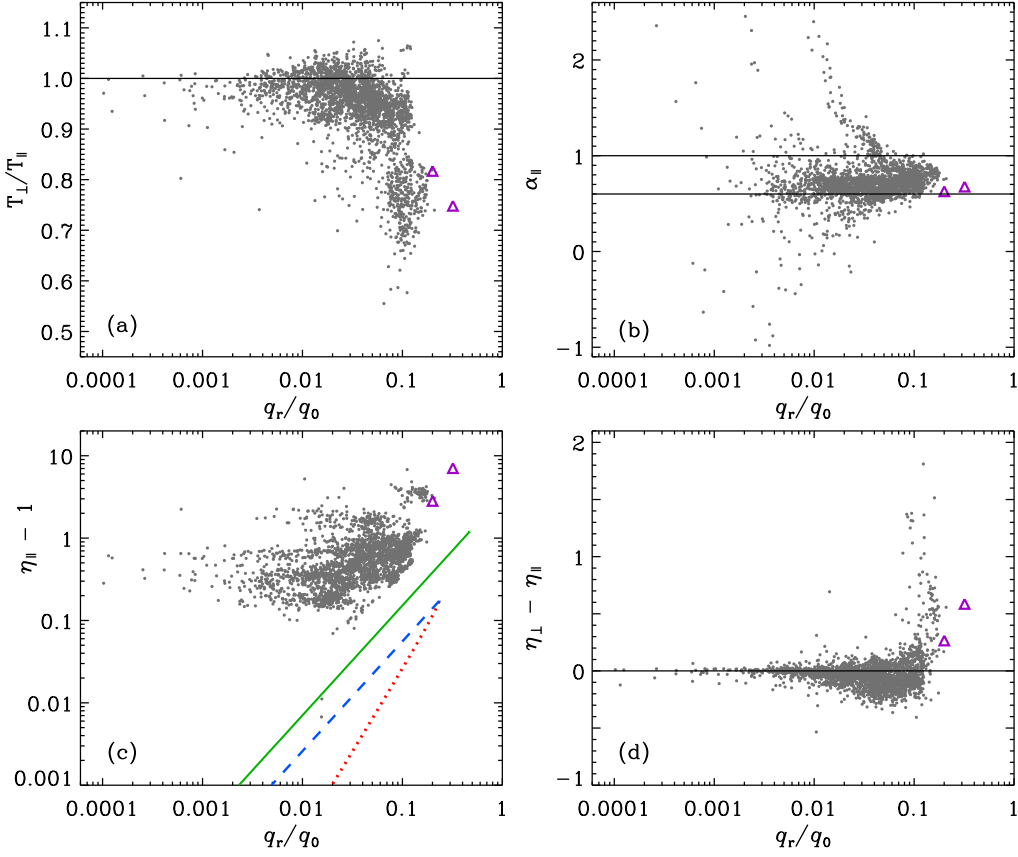}
\caption{Dimensionless moment quantities for measured VDFs of
\citeA{Wi19a} (gray dots) and \citeA{Sk21} (purple triangles)
plotted versus normalized heat flux $q_r / q_0$.
(a) Anisotropy ratio $T_{\perp}/T_{\parallel}$, with the isotropic
value of 1 highlighted by a horizontal line.
(b) Third moment partition fraction $\alpha_{\parallel}$, with
analytic values (0.6, 1) highlighted by lines.
(c) Fourth moment quantity ($\eta_{\parallel}-1$), with analytic
results shown with curves of same colors/styles of Figure \ref{fig02}.
(d) Fourth moment difference ($\eta_{\perp}-\eta_{\parallel}$).}
\label{fig03}
\end{figure}

Figure \ref{fig03}a shows that the electron anisotropy ratio
$T_{\perp}/T_{\parallel} = {\cal P}_{\theta\theta} / {\cal P}_{rr}$
is generally near unity for solar-wind electrons at 1~AU.
The median of the distribution of the \citeA{Wi19a} values is 0.959,
with $\pm 1$ standard deviation bounds at values of 0.87 and 1.01.
The two \citeA{Sk21} models are on the more anisotropic side, with
$T_{\perp}/T_{\parallel} =$ 0.748 and 0.817.
Despite these anisotropy ratios not being exactly equal to one,
they seem sufficiently close that the results for the third-order
and fourth-order moments ought to be at least roughly comparable to
the isotropized analytic models of
Sections \ref{sec:vdf:grad}--\ref{sec:vdf:whealton}.

Figure \ref{fig03}b shows how the third-order partition fraction
$\alpha_{\parallel}$ does not seem to depend strongly on the value of
the heat flux.
The larger spread at lower values of $q_r/q_0$ may be the result of
fitting uncertainties when the skewness is weak, but it is notable
that the other panels of Figure \ref{fig03} do not show this kind of
spread at low $q_r/q_0$.
The median of the distribution of plotted values of $\alpha_{\parallel}$
is 0.704, with $\pm 1$ standard deviation limits between 0.59 and 0.83.
Note that roughly three-quarters of the plotted values (i.e., 2,175
out of 2,840) fall in between the analytic values corresponding to
the Grad-expansion VDFs ($\alpha_{\parallel} = 0.6$) and the
other analytic VDFs ($\alpha_{\parallel} = 1$).
The two \citeA{Sk21} cases have $\alpha_{\parallel} = 0.628$ and 0.677.

Figure \ref{fig03}c shows the fourth-order moment quantity
$\eta_{\parallel}$ alongside the three sets of analytic trends from
Figure \ref{fig02}.
For both sets of measured data, the full range of $\eta_{\parallel}$
values extends from 1.007 to 8.05.
There is a hint of a trend of positive correlation with
$q_r/q_0$ (like in the analytic models), but it is not a strong trend.
It may then be reasonable to just adopt the median value
($\eta_{\parallel} = 1.492$) for the data in future modeling efforts.
Note that the existence of $\kappa$ tails in the measured VDFs
gives rise to larger values of the fourth moment than were found in the
analytic VDF models discussed above.
This was anticipated in equation (\ref{eq:kappa4th}).

Figure \ref{fig03}d shows that $\eta_{\perp}$ tends to remain close
to $\eta_{\parallel}$, with a median value of their difference being
only $(\eta_{\perp} - \eta_{\parallel}) = -0.033$.
Note, however, that the median of the distribution of $\eta_{\perp}$ 
values themselves is 1.444, which corresponds to a slightly different
value of the difference when the two medians are subtracted.
One can also see that, even though these median differences are negative,
the cases of strongest heat flux---including the two VDFs described
by \citeA{Sk21}---tend to have positive values.
Thus, in some situations it may be appropriate to just adopt
$\eta_{\perp} = \eta_{\parallel}$, which simplifies the equations
for $q_r$.

\section{Applicability to the Solar Wind}
\label{sec:wind}

Although the results of Sections \ref{sec:cons} and \ref{sec:vdf}
are interesting on their own, we would like to also begin exploring
solutions for $q_r$ with ambient plasma conditions relevant to
the observed solar wind.
However, the solutions presented below should be considered only as
example cases that scratch the surface of how
equations (\ref{eq:dqdt_v3})--(\ref{eq:qden}) can be applied once
one adopts particular values for the VDF-specific third-order
and fourth-order constants.
For example, several eight-moment and gyrotropic models of the
solar wind \cite{OL96,OL99,S20} used the values from the
Grad-type \cite{Sc77} expansion:
\begin{equation}
  \left\{ \alpha_{\parallel}, \alpha_{\perp},
  \eta_{\parallel}, \eta_{\perp} \right\} \, = \,
  \left\{ 0.6, 0.4, 1, 1 \right\}
  \label{eq:param_grad} 
\end{equation}
\cite<see also>[]{BS93,Li99,LS01,KD09}.
We will also explore the implications of using the median values
derived from {\em Wind} data and discussed in
Section \ref{sec:vdf:data}:
\begin{equation}
  \left\{ \alpha_{\parallel}, \alpha_{\perp},
  \eta_{\parallel}, \eta_{\perp} \right\} \, = \,
  \left\{ 0.704, 0.296, 1.492, 1.444 \right\}  \,\, .
  \label{eq:param_median}
\end{equation}
Section \ref{sec:wind:scaling} examines some potentially useful
closed-form limiting cases,
Section \ref{sec:wind:num} presents numerical solutions, and
Section \ref{sec:wind:ext} discusses how things change if we
include external source/sink terms in the conservation equations.

\subsection{Analytic Scaling Relations}
\label{sec:wind:scaling}

Although the specification of $q_r$ in
equations (\ref{eq:qnumden})--(\ref{eq:qden}) is the most general
result of this paper, we find it useful to work with a simplified
version that can be applied in much of the solar wind.
Sufficiently far from the Sun, the solar wind has accelerated to a
nearly constant speed (so that $\partial u_r / \partial r \approx 0$)
and the expansion is close to spherically symmetric (so that
$A \propto r^2$ and $L_{\rm A} \approx r/2$).
In that case, mass-flux conservation demands that $n \propto r^{-2}$
and we can parameterize the radial dependence of electron temperature
as $T \propto r^{-\delta}$.
This gives
\begin{equation}
  q_r \, \approx \,
  \frac{p u_r^2 (4 - 3\delta) + 5 p w^2 ( \eta_{\perp} +
  \delta \eta_{\parallel} - 1 - \delta/2 )}
  {4\alpha_{\perp} u_r + 2 \nu_{\rm eff} r} \,\, .
  \label{eq:qouter}
\end{equation}
In the weak-skewness limit of
$\eta_{\parallel} \approx \eta_{\perp} \approx 1$, the second
expression in parentheses above reduces to $\delta / 2$, and
Section \ref{sec:cons:coll} showed how $\nu_{\rm eff}$ can be
defined with a normalizing constant $\xi = 0.8325$.
However, for other values of $\eta_{\parallel}$ and $\eta_{\perp}$,
one would have to use
\begin{equation}
  \xi \, = \, 0.8325 \left( 
  \frac{\eta_{\perp} + \delta \eta_{\parallel} - 1 - \delta/2}
  {\delta/2} \right)
\end{equation}
in order to ensure \citeA{SH53} conductivity in the limit of rapid
collisions for a static ($u_r = 0$) plasma.

If we assume that collisions are weak and the electron temperature is
roughly constant with distance (i.e., $\delta \approx 0$), together
with the approximation $\eta_{\perp} \approx 1$, then
equation (\ref{eq:qouter}) reduces to
\begin{equation}
  q_r \, \approx \, \frac{p u_r}{\alpha_{\perp}} \,\, ,
\end{equation}
which is essentially the free-streaming limit derived by \citeA{H74,H76}:
\begin{equation}
  q_r \, \approx \, \frac{3}{2} \alpha_{\rm H} u_r n k_{\rm B} T
  \, = \, \frac{3}{2} \alpha_{\rm H} p u_r \,\, .
\end{equation}
Typical values of $\alpha_{\perp}$ of 0.3--0.4 give a predicted range
for Hollweg's constant $\alpha_{\rm H}$ of 1.7--2.2.
This falls within the somewhat broader range of likely values
($1 < \alpha_{\rm H} < 4$) discussed by \citeA{H74,H76}.

As shown below, the limit of $\delta \approx 0$ is not often seen in
the solar wind.
For realistic temperature gradients and outflow speeds, the first
($u_r^2$ dependent) term in the numerator of equation (\ref{eq:qouter})
tends to be much smaller in magnitude than the second ($w^2$ dependent)
term in much of the heliosphere.
In that case, the limit of weak collisions would provide a heat flux
that scales as
\begin{equation}
  q_r \, \propto \, \frac{p w^2}{u_r} \,\, ,
  \label{eq:qranti}
\end{equation}
with a dimensionless multiplicative constant (depending on
$\alpha_{\perp}$, $\delta$, $\eta_{\parallel}$, and $\eta_{\perp}$)
left unwritten.
This type of collisionless heat-flux relation does not resemble either
the free-streaming limit ($q_r \propto p u_r$) or the
saturated limit ($q_r \propto pw$).
Note that there is no combination of terms in the numerator and
denominator of equation (\ref{eq:qouter}) that yields the well-studied
saturated limit of $q_r \approx q_0$.
\citeA{Ha21} found an approximate anticorrelation between the heat
flux and the solar wind speed in data from
{\em Parker Solar Probe} ({\em{PSP}}),
which may support scaling relations like equation (\ref{eq:qranti}).

Equation (\ref{eq:qouter}) resembles a frequently-used way of
specifying a gradual transition from collisional to collisionless
heat flux in solar-wind simulations.
For example, \citeA{CvB07} followed \citeA{Wa93} and others to adopt
a form that was given as
\begin{equation}
  q_r \, = \, \frac{\nu_{\rm exp} q_{\rm FS} + \nu_{\rm coll} q_{\rm SH}}
  {\nu_{\rm exp} + \nu_{\rm coll}}  \,\, ,
  \label{eq:qzephyr1}
\end{equation}
where $q_{\rm SH}$ is the Spitzer-H\"{a}rm heat flux, $q_{\rm FS}$ is an
estimate of the free-streaming (collisionless) heat flux taken from
\citeA{H74,H76}, $\nu_{\rm coll} \approx \nu_{\rm eff}$ is an electron
self-collision frequency, and $\nu_{\rm exp}$ is a local rate of
solar-wind expansion.
When these terms are translated into the notation of this paper, this
expression becomes
\begin{equation}
  q_r \, \approx \, \frac{3 \alpha_{\rm H} p u_r^2 + 1.59 \delta p w^2}
  {2 u_r + 1.28 \nu_{\rm eff} r} \,\, ,
  \label{eq:qzephyr2}
\end{equation}
which closely resembles equation (\ref{eq:qouter}) above.
Heat-flux formulations like equation (\ref{eq:qzephyr1}) are intended
be strongly collisional near the Sun (where
$\nu_{\rm coll} \gg \nu_{\rm exp}$) and collisionless far from the Sun
(where $\nu_{\rm coll} \ll \nu_{\rm exp}$).
However, these expected radial variations in the relative values of
$\nu_{\rm coll}$ and $\nu_{\rm exp}$ may not be so straightforward.
In the denominator of equations (\ref{eq:qouter}) and (\ref{eq:qzephyr2}),
the $\nu_{\rm eff} r$ term may either increase or decrease with
increasing $r$, depending on the value of $\delta$.
Specifically, $\nu_{\rm eff} r \propto r^{(3\delta - 2)/2}$, so it
only grows weaker with distance when $\delta < 2/3$.

\subsection{Representative Numerical Models}
\label{sec:wind:num}

To illustrate the behavior of the modeled heat flux as a function
of heliocentric distance, we evaluated equations
(\ref{eq:qnumden})--(\ref{eq:qden}) numerically using tabulated
models of solar-wind plasma properties.
The outflow speed $u_r$, number density $n$, and flux-tube
area expansion factor $A$ were taken from a semi-empirical model of the
time-steady fast solar wind \cite{CvB12}.
For the electron temperature $T$, we wanted to vary the radial
exponent $\delta$ as a free parameter, so we used the analytic model
described further in the Appendix.
In all cases shown below, the following parameters of that model were
fixed: $T_{\rm max} = 1.35$~MK, $\psi = 4.5$, and the latter is
consistent with a dimensionless height of the temperature maximum of
$x_{\rm max} = 1.842$.
These values were chosen to match a recent set of collected observations
of the electron temperature above polar coronal holes \cite{Cr20}.

\begin{figure}[!t]
\centering%
\noindent\includegraphics[width=0.95\textwidth]{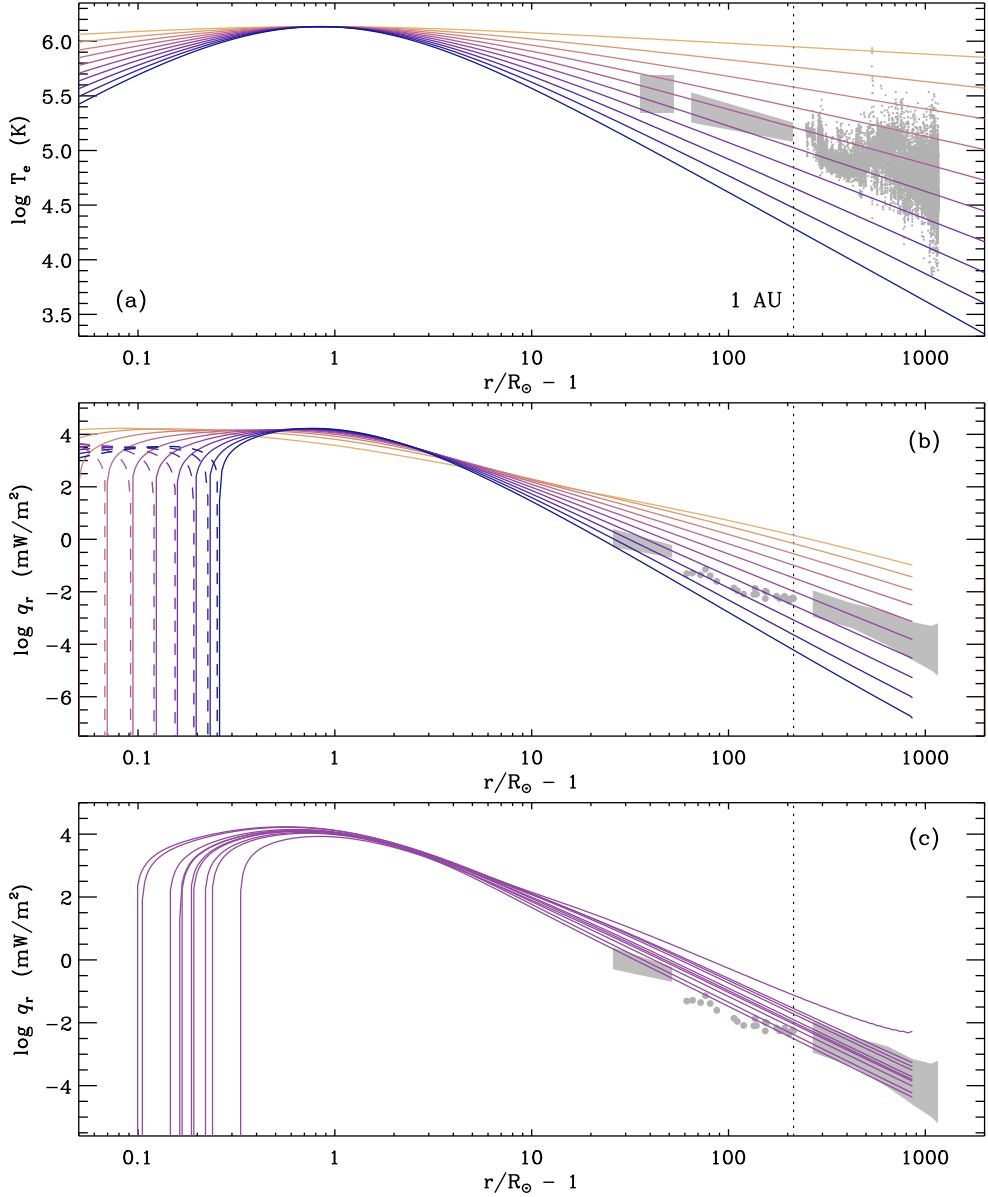}
\caption{(a) Radial dependence of electron temperature for
analytic models (solid curves) and in~situ data (gray regions).
Curve colors go from gold ($\delta = 0.1$) to violet ($\delta = 1$).
(b) Computed values of $q_r$ for temperatures shown in panel (a), using
equation (\ref{eq:param_median}), also compared with in~situ data.
Solid curves show $q_r > 0$, and dashed curves show $|q_r|$ when $q_r < 0$.
(c) Computed values of $q_r$ for the $\delta = 0.6$ model and a
selection of higher-moment constants from the \citeA{Wi19a} database,
with negative values not shown.}
\label{fig04}
\end{figure}

Figure \ref{fig04}a shows a range of electron temperature models,
with $\delta$ varied between 0.1 and 1 in steps of 0.1.
Measured values from the solar wind come from
{\em PSP} \cite{Ha20},
{\em Helios} \cite{Mk20}, and
{\em Ulysses} \cite{Ba92,Go96}, with the processing of
the latter two sources discussed in more detail by \citeA{Cr09}.
For these data, the best-fitting value of $\delta$ seems to be about 0.6.
Note that at 1~AU, an electron temperature of $\sim$10$^5$~K
corresponds to $w \approx$~1,750 km~s$^{-1}$, which is
significantly larger than typical solar-wind speeds of
$u_r \approx 300$--700 km~s$^{-1}$.

Figure \ref{fig04}b shows corresponding solutions to
equations (\ref{eq:qnumden})--(\ref{eq:qden}) for the median
parameter values given in equation (\ref{eq:param_median}).
To compute the collision rate $\nu_{\rm eff}$, we needed to evaluate
the Coulomb logarithm,
\begin{equation}
  \ln \Lambda \, = \, 23.2 + \frac{3}{2} \ln \left(
  \frac{T}{10^{6} \, \mbox{K}} \right) - \frac{1}{2} \ln \left(
  \frac{n}{10^{6} \, \mbox{cm}^{-3}} \right)  \,\, .
\end{equation}
Also shown are measured values from {\em PSP} \cite{Ha21},
{\em Helios} \cite{Pi90}, and {\em Ulysses} \cite{Se01}, and it
is clear that the best-fitting temperature exponent of $\delta = 0.6$
also appears to produce heat fluxes that agree somewhat well with
the data.
Note that the radial distance where $q_r = 0$ occurs at slightly
different places in the low corona, depending on the value of $\delta$,
and it never coincides exactly with the location where
$\partial T / \partial r = 0$.
Below, we discuss comparisons between these numerical calculations
and the limiting cases of Spitzer-H\"{a}rm heat flux ($q_{\rm SH}$)
and saturated heat flux ($q_0$).

In Figure \ref{fig04}c, we fix $\delta$ at its best-fitting value of
0.6 and explore what happens when the higher-moment parameters
($\alpha_{\parallel}$, $\alpha_{\perp}$,
$\eta_{\parallel}$, $\eta_{\perp}$) are varied.
The curves in Figure \ref{fig04}c represent 10 different selections
of parameters from the list of 2,840 cases in the \citeA{Wi19a} database.
Generally, larger values of $\alpha_{\parallel}$ (i.e., smaller values
of $\alpha_{\perp}$) corresponded to larger values of $q_r$ at 1~AU,
but no definitive correlations were seen with the $\eta_{\parallel}$
or $\eta_{\perp}$ parameters.
Note that even when holding $\delta$ fixed, the radial distance where
$q_r = 0$ can vary, depending on the values of the higher-moment parameters.
It is clear that the electron heat flux in interplanetary space
depends on both macroscopic properties of the expansion (i.e., $\delta$)
and microscopic properties of the skewed electron VDF.

\begin{figure}[!t]
\centering%
\noindent\includegraphics[width=0.80\textwidth]{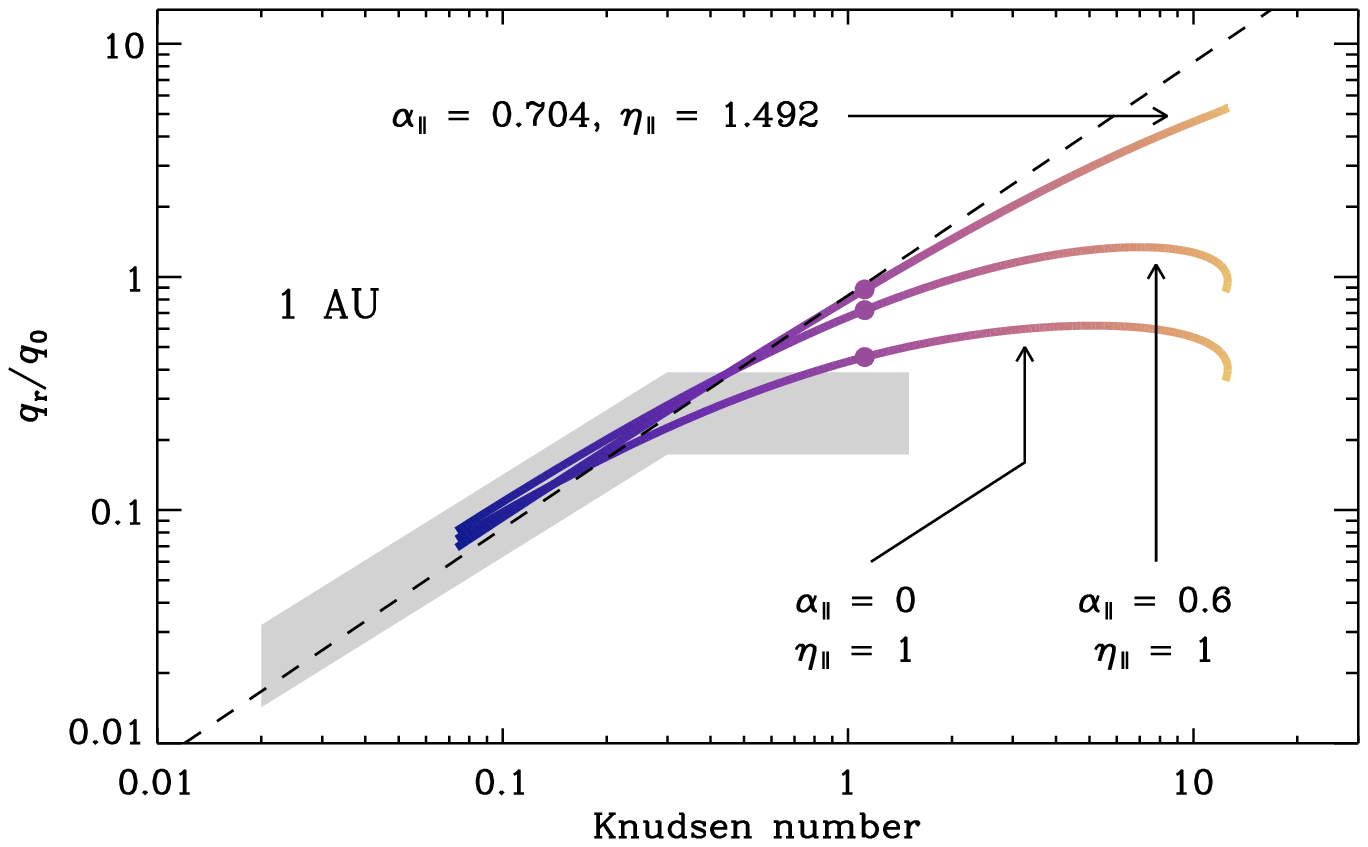}
\caption{Normalized heat flux ratio $q_r/q_0$ at 1~AU plotted versus
Knudsen number, for three sets of higher-moment constants (see text).
Solid curves denote values modeled with a fine grid of temperature
gradients, from $\delta = 0.1$ (gold) to $\delta = 1$ (violet),
with the best-fitting value of $\delta = 0.6$ highlighted with circles.
Also shown: $q_{\rm SH}/q_0$ (black dashed curve) and an approximate
range of measured data from {\em{Wind}} (\citeA{Ba13}; gray region).}
\label{fig05}
\end{figure}

Figure \ref{fig05} shows modeled values of the ratio $q_r/q_0$ at a
fixed distance of $r=1$~AU, plotted versus the dimensionless Knudsen
number that is defined as
\begin{equation}
  \mbox{Kn} \, = \, \frac{w}{\nu_{\rm eff}} \left|
  \frac{\partial \ln T}{\partial r} \right| \,\, .
\end{equation}
Here, we use the fixed value of $\xi = 0.8325$ in the definition
for $\nu_{\rm eff}$ in order to have a single consistent scale.
Traditionally, values of $\mbox{Kn} \ll 1$ denote strongly
collisional regimes and values of $\mbox{Kn} > 1$ denote
collisionless regimes.
Note that the range of plotted $\mbox{Kn}$ values is slightly larger
than the typical observed range \cite{Sa03,Ba13}.
This is probably because the bulk outflow speed and density in our
models were specified for the fast solar wind, which tends to be less
collisional than the slow solar wind.

The curves in Figure \ref{fig05} represent finer grids of $\delta$
values than were shown in Figure \ref{fig04}, and we also selected
three different sets of higher-moment constants: those from
equations (\ref{eq:param_grad}) and (\ref{eq:param_median}), and one
with $\alpha_{\parallel}=0$ and $\eta_{\parallel}=\eta_{\perp}=1$
to illustrate the lowest heat fluxes that are likely to be obtained
from this model.
In this kind of diagram, the Spitzer-H\"{a}rm limit is
a straight line, with $|q_{\rm SH}|/q_0$ varying linearly with $\mbox{Kn}$.
The order-unity normalization factor between these two quantities
is often given as 1.07, but its precise value depends on details of
how the Coulomb collision rates and mean-free paths are defined.
Using the definitions given above, that factor here is 0.833.

Although the model curves shown in Figure \ref{fig05} do not agree
perfectly with the data, it is interesting that they still show a
similar trend of dipping below the Spitzer-H\"{a}rm value at the
largest Knudsen numbers.
We also note, following \citeA{Ln14}, that just because a point may
happen to be near the linear $q_{\rm SH}/q_0$ relationship on this
diagram, it does not mean that the corresponding heat flux is
well-described by classical \citeA{SH53} conductivity.
In reality, we expect that kinetic instability thresholds (that are
not yet included in our model) must be responsible for limiting the
observed heat fluxes to values below $q_r/q_0 \approx 0.3$--0.4;
see Section \ref{sec:conc} below.

\subsection{Effects of External Sources and Sinks}
\label{sec:wind:ext}

The conservation equations for electron VDF moments do not
exist in a vacuum.
As mentioned in Section \ref{sec:cons:mom}, there are several ways
that other particle species and external forces can give rise to
additional terms in these equations.
For example, in the momentum equation, electrons feel the effects
of gravity (inward), MHD-wave pressure (outward), and the Lorentz
force associated with a charge-separation electric field
(usually inward for electrons).
In the thermal energy equation, there can be multiple competing sources
of heating and cooling from collisions, radiation, or wave-particle
interactions.

In order to explore the importance of these source/sink terms to the
determination of the electron heat flux, we can add an arbitrary
vector momentum density ${\bf M}$ to the right-hand side of
equation (\ref{eq:mom_orig}) and a scalar heating term $Q$ to the
right-hand side of equation (\ref{eq:dpdt_orig}).
If we also rewrite the momentum source as ${\bf M} = mn{\bf a}$,
where ${\bf a}$ is an acceleration, then these terms can be carried
through to the analytic solution given by
equations (\ref{eq:qnumden})--(\ref{eq:qden}).
In that solution, we would replace $q_{\rm num}$ by
$\tilde{q}_{\rm num}$, and the latter is defined as
\begin{equation}
  \tilde{q}_{\rm num} \, = \, q_{\rm num} 
  - \frac{2}{5} u_r Q - p a_r \,\, .
  \label{eq:external}
\end{equation}
Note that both new terms are negative, since the presence of local
sources of electron momentum or energy would reduce the need for
radial heat flux to have to transport energy out to maintain a given
steady state.

There have been several estimates of the magnitude of the electron
heating rate $Q$ in the solar wind.
\citeA{Va07}, \citeA{Cr09}, and \citeA{St09} found values between
roughly $10^{-17}$ and $10^{-16}$ W~m$^{-3}$ at 1~AU.
This range would produce terms in equation (\ref{eq:external}) of order
$10^{-12}$ to $10^{-11}$ W~m$^{-2}$~s$^{-1}$.
However, the $\delta = 0.6$ model of Figure \ref{fig04}b exhibited
a value of $q_{\rm num} \approx 10^{-10}$ W~m$^{-2}$~s$^{-1}$
at 1~AU.
Thus, these heating terms may only be 1\% to 10\% corrections to
the electron heat flux computed without them.
In addition, \citeA{Sv15} concluded that the uncertainties inherent
in many observational determinations of the electron heating rate may
be high enough that we are not even sure whether $Q$ is positive or
negative at 1~AU.
Thus, we defer further discussion of these source/sink terms
(including the $a_r$ term) to future work.

\section{Discussion and Conclusions}
\label{sec:conc}

It is a long-standing goal of heliophysics to produce self-consistent
models of coronal heating, solar wind acceleration, and the multi-scale
structure of the heliosphere.
If such models are to employ fluid-based conservation equations for
the moments of electron, proton, and heavy-ion VDFs, it is necessary
for those equations to be free from overly limiting or inconsistent
assumptions about the kinetic microphysics.
For example, we have known for decades that it is inappropriate to use
the classical Spitzer-H\"{a}rm heat flux in regions with infrequent
collisions.
There are also major problems with using polynomial VDF expansions
that provide collisionless expressions for heat flux at the expense of
creating large regions of unphysically negative phase-space density.
Thus, in this paper we have attempted to navigate this minefield by
investigating multiple ways of expressing heat-flux-carrying skewness
in a model electron VDF, and then exploring how each kind of skewed
distribution affects the higher-moment conservation equations for
the heat flux itself.

Much of the effort in this paper was devoted to computing values for
the third-moment ($\alpha_{\parallel}$, $\alpha_{\perp}$) and
fourth-moment ($\eta_{\parallel}$, $\eta_{\perp}$) partition fractions
needed to specify all terms in the heat-flux conservation equation.
Both idealized analytic VDFs
(Sections \ref{sec:vdf:dimax}--\ref{sec:vdf:whealton})
and fits to measured solar-wind electron data (Section \ref{sec:vdf:data})
showed roughly similar ranges of likely values for the partition fractions.
Further work is definitely needed to produce optimally customized sets of
parameters for use in models of specific regions of the heliosphere.
However, it is interesting that the higher-moment parameters of the flawed
Grad-type expansion technique (Section \ref{sec:vdf:grad}) were not
{\em too} far removed from the other values computed for
positive-definite VDFs.
This suggests that some results of eight-moment and gyrotropic
models of the solar wind may yet have value despite their being
built on somewhat shaky foundations.

To improve our current descriptions of electron heat flux, it is
clear that more kinetic theory is needed.
In recent years, insights from collisionless exospheric approaches to
the solar wind have begun to be appreciated more by the fluid-based
modeling community \cite<see, e.g.,>[]{Ec11,VB20}.
There may also be specific kinetic connections between the solar
atmosphere and heliosphere that depend on the physics of coronal heating;
i.e., nanoflare-generated beams that can affect the electron strahl
and halo in the heliosphere \cite{Che14}.
Perhaps most importantly, the effects of kinetic instabilities must also
be included when attempting to determine the electron heat flux in
interplanetary space \cite{Fo70,Pe73,Ga75,Se94,Sh18,To19,In20,Lo20,Mi20}.

It is possible that new analytic forms of skewed VDFs can be constructed
that represent observed heat-flux trends better than the ones discussed
in this paper.
Parameterizations that treat the electron strahl as a narrow beam
in velocity space \cite<e.g.,>[]{Ho18}
or as a truncated bi-kappa function \cite{Sv09}
may be particularly useful.
Alternative moment-closure ideas, such as maximum entropy constraints
\cite{Lv96,GM09}, perturbative expansions that go to higher orders
\cite{Mi65,ST03,GC08}, or spatially nonlocal treatments that scale
with the mean-free path \cite{Lu83,Cn96} could be helpful as well.
Lastly, we note that collisionless modifications to the heat flux
carried by electrons may be important in astrophysical contexts far
beyond our heliosphere \cite<see, e.g.,>[]{CM77,NM01,Vo11}.

\appendix
\section{Analytic RTV-like Temperature Laws for the Solar Wind}

For time-steady heating, the solar corona usually exhibits a time-steady
solution for temperature and density as a function of spatial position.
\citeA{RTV}, hereafter RTV, produced analytic and semi-analytic solutions
for closed coronal loops \cite<see also>[]{Se81,As01}.
\citeA{Ma10} showed that, for a specific way of parameterizing
the heating rate, the dependence of temperature on position
can be written using the inverse of an incomplete beta function.
However, for the open field lines that connect the Sun to the heliosphere,
there do not seem to be closed-form solutions of this kind.
\citeA{SM03} laid out some necessary ingredients for such a calculation,
but no analytic solutions for $T(r)$ were found.
In this appendix we reproduce a derivation, originally given by \citeA{S20},
for a closed-form solution to a simplified version of the thermal energy
conservation equation in an open-field geometry.

Normally, the solution of the thermal energy equation in the corona is
a balance between terms describing direct heating, thermal conduction,
radiative losses, and enthalpy transport due to flows.
\citeA{Br19} found that accounting for only a subset of these terms
can still produce reasonable solutions at times.
Let us assume, for the low corona and upper transition region, that
the radiative losses are negligible (because the electron density
begins to decrease rapidly with increasing height) and that the enthalpy
transport terms are also negligible (because the solar wind has not
yet accelerated to high speeds).
In that case, the time-steady energy balance contains only terms
describing direct heating and thermal conduction:
\begin{equation}
  Q_{\rm heat} + Q_{\rm cond} \, = \, 0 \,\, ,
  \label{eq:energybalance}
\end{equation}
where the $Q$ terms are volumetric energy deposition rates in the same
units as the terms in equation (\ref{eq:dpdt_v2}).
Describing the heating using a power-law function of heliocentric
distance $r$ \cite<see, e.g.,>[]{Hu97,Cr09}, we can write
\begin{equation}
  Q_{\rm heat} \, = \,
  Q_{\odot} \left( \frac{R_{\odot}}{r} \right)^{\psi}
\end{equation}
where $Q_{\odot}$ is a normalizing constant that sets the rate of
heating at the base of the corona, and $\psi$ describes its power-law
rate of radial decrease.
For simplicity, let us describe the thermal conduction using the
Spitzer-H\"{a}rm limit in spherical geometry, with
\begin{equation}
  Q_{\rm cond} \, = \, \frac{\kappa_0}{r^2} \frac{d}{dr}
  \left( r^2 T^{5/2} \frac{dT}{dr} \right) \,\, .
  \label{eq:Qconddef}
\end{equation}
If we make the following change of variables,
\begin{equation}
  x \, = \, \frac{r}{R_{\odot}}
  \,\,\, , \,\,\,\,\,\,
  y \, = \, \left( \frac{T}{T_{\rm max}} \right)^{7/2}
  \label{eq:xydefs}
\end{equation}
then equation (\ref{eq:energybalance}) can be written as
\begin{equation}
  \frac{\zeta}{x^{\psi}} + \frac{1}{x^2} \frac{d}{dx}
  \left( x^2 \frac{dy}{dx} \right) \, = \, 0 \,\, ,
  \label{eq:zetaODE}
\end{equation}
where the dimensionless constant $\zeta$ is given by
\begin{equation}
  \zeta \, = \,
  \frac{7 R_{\odot}^2 Q_{\odot}}{2 \kappa_0 T_{\rm max}^{7/2}}
  \,\, .
  \label{eq:zetadef}
\end{equation}

Note that we have introduced an additional unknown parameter into the
system: the peak coronal temperature $T_{\rm max}$.
Following \citeA{RTV}, we realize that {\em overspecifying} the boundary
conditions can lead to the ability to solve for not only this new
parameter, but also for the radial distance $x_{\rm max}$ at which
$T = T_{\rm max}$.
Equation (\ref{eq:zetaODE}) is a second-order differential equation,
which requires specifying only two boundary conditions on the desired
solution for $y(x)$.
However, let us specify four such conditions:
\begin{equation}
  y(1) = 0
  \,\,\, , \,\,\,\,\,\,
  y(x_{\rm max}) = 1
  \,\,\, , \,\,\,\,\,\,
  y'(x_{\rm max}) = 0
  \,\,\, , \,\,\,\,\,\,
  y(\infty) = 0
  \,\, .
\end{equation}
In other words, we assume the chromospheric lower boundary at $x=1$
is sufficiently ``cold'' such that $T \ll T_{\rm max}$, and that the
temperature also asymptotes to zero at large enough distances from the Sun.
Also, we assume that the height $x_{\rm max}$ at which the temperature
reaches its peak is a well-behaved local maximum where the first
derivative of $y$ is equal to zero.

Depending on the value of the exponent $n$, there are three possible
closed-form solutions to equation (\ref{eq:zetaODE}).
There are special cases for $\psi=2$ and $\psi=3$, as well as a
general solution,
\begin{equation}
  y(x) \, = \, \frac{C_1}{x} + C_2
  - \frac{\zeta x^{2-\psi}}{\psi^2 - 5\psi + 6}
\end{equation}
where $C_1$ and $C_2$ are constants that must be fixed by applying
the boundary conditions, and this expression does not hold for either
$\psi=2$ or $\psi=3$.
The boundary conditions on $y(1)$ and $y(\infty)$ demand that
\begin{equation}
  C_1 \, = \, \frac{\zeta}{\psi^2 - 5\psi + 6}
  \,\,\, , \,\,\,\,\,\,
  C_2 \, = \, 0 
\end{equation}
and the other boundary conditions help to determine
\begin{equation}
  x_{\rm max} \, = \, (\psi-2)^{1/(\psi-3)}
  \,\,\, , \,\,\,\,\,\,
  \zeta \, = \, (\psi-2)^{(2\psi-5)/(\psi-3)} \,\, .
\end{equation}
This yields a closed-form solution given by
\begin{equation}
  y(x) \, = \, \frac{(\psi-2)^{(\psi-2)/(\psi-3)}}{\psi-3}
  \left( \frac{1}{x} - \frac{1}{x^{\psi-2}} \right) \,\, .
\end{equation}
As mentioned above, the overspecification of boundary conditions
also allows a numerical value of $\zeta$ to be computed, which provides
a scaling law similar to that of \citeA{RTV}.
The relationship between peak temperature and the surface heating
rate (i.e., $T_{\rm max} \propto Q_{\odot}^{2/7}$) is identical to that
of RTV, which conveys that the general nature of this scaling law
does not depend on the inclusion or neglect of radiative losses.

The above solution for $y(x)$ exhibits a maximum value at a radial
distance that depends inversely on the exponent $\psi$.
A more rapidly dropping heating rate corresponds naturally to a
temperature that reaches its peak value closer to the Sun. 
At large distances from the Sun, the above solution behaves as
$y \propto 1/x$, which implies $T \propto r^{-2/7}$ as originally
described by \citeA{Ch57} for a conduction-dominated corona.
Our semi-empirical description of the solar wind in
Section \ref{sec:wind:num} makes use of a slight modification to
equation (\ref{eq:xydefs}), which is
\begin{equation}
  y \, = \, \left( \frac{T}{T_{\rm max}} \right)^{1/\delta} \,\, .
\end{equation}
This allows the large-scale temperature dropoff to be more
freely adjustable as $T \propto r^{-\delta}$.
However, it also means that the solution no longer exactly satisfies
equations (\ref{eq:energybalance})--(\ref{eq:Qconddef}).
Thus, this $\delta$-modification should be viewed only as a convenient
parameterization and not reflective of real physics.

\section*{Data Availability Statement}

Model data generated for this paper, including some ``data behind the
figures'' and other supplemental information, have been published to the
Zenodo data repository and are available in \citeA{Cr21}.
Other observational data used in this paper have all been obtained
from previously published sources
\cite<see, e.g.,>[]{Pi90,Se01,Cr09,Wi19a,Ha20,Mk20,Ha21,Sk21}.

\acknowledgments
The authors gratefully acknowledge Jack Scudder
for many valuable discussions.
The authors are also grateful to the anonymous referee for
constructive suggestions that have improved this paper.
This work was supported by the National Science Foundation (NSF)
Graduate Research Fellowship Program under grant 1650115.
Additional support came from the National Aeronautics and Space
Administration (NASA), under grants {NNX\-16\-AG87G} and
{80\-NSSC\-20K\-1319}, and from the NSF under grant 1613207.
This research made extensive use of NASA's Astrophysics Data System (ADS).
The data on which this article is based are available in \citeA{Cr21}.

\end{document}